\documentclass[aps, prl, twocolumn, floatfix, 10pt]{revtex4-1}

\usepackage{amsmath}
\usepackage{amssymb}
\usepackage{graphicx}
\usepackage{wasysym}
\usepackage{hyperref}

\newcommand{\showMLMR}[1]{}%{#1}

\begin{document}

\title{Cross-Polarized Microwave Surface-State Anti-Resonance}

\author{Ian Appelbaum}
\altaffiliation{appelbaum@physics.umd.edu}
\affiliation{Department of Physics and Center for Nanophysics and Advanced Materials, University of Maryland, College Park, Maryland 20742, USA}

\begin{abstract}
We propose a polarization-sensitive measurement of microwave electromagnetic resonances in a static magnetic field to detect the metallic surface state of a bulk insulator. A quantitative model is used to demonstrate that a unique, unambiguous signature of the dissipative but conducting surface can be seen in the orthogonally polarized transmission spectra.   
\end{abstract}

\maketitle

\emph{Introduction--} Discerning the role of bulk and surface-state contributions to conductivity has become an important experimental task in the context of identifying possible 3-dimensional topological insulators (TIs).\cite{Hasan_RMP2010} Most arguments claiming transport-based observation of these states (whether ``trivial'' or ``topological'') appeal only to circumstantial evidence, such as measurement of residual conductivity, geometrical variation, etc. A clear, unambiguous signature of such a conductive state is desired.

Here, we propose to detect the presence of a conducting surface state by utilizing it as a metallic resonator coupling cross-polarized wide-bandwidth microwave waveguides\cite{Portis_JAP1958, Moore_RSI1973}. A longitudinal magnetic field $B_z$ introduces non-zero off-diagonal conductivity tensor components\cite{Suhl_PR1952} that provides polarization rotation into an orthogonal linear orientation, similar to the role of cavity-wall conduction in microwave Hall measurements\cite{Trukhan_REEP1966, Fletcher_JPhysE1976, Eley_JPhysE1983}. The cross-polarized transmission spectrum then carries a unique signature distinct from bulk Faraday effect in the regime $B_z \mu\lesssim 1$, where $\mu$ is planar mobility of the carriers in the surface state. 

Polarization rotation (i.e. surface Kerr or bulk Faraday rotation) from surface states of topological insulators in a magnetic field has been discussed several times previously in the literature.\cite{Aguilar_PRL2011, Maciejko_PRL2010,Shuvaev_PRL2011, Tkachov_PRB2011, Tse_PRL2010, Tse_PRB2011} However, the coherent superposition of electromagnetic fields resulting in periodic transmission features has to our knowledge not been examined or exploited experimentally. To provide quantitative insight on this problem, we analyze a simplified 1-d slab geometry using a coherent transfer matrix approach, and derive analytic expressions defining the transitions between three distinct regimes.

\emph{Model--} Consider an infinite planar slab dielectric of thickness $L$ with internal index of refraction $n=\sqrt{\epsilon}$, where $\epsilon$ is the relative dielectric permittivity. This slab also has a 2-dimensional conductor on the interfaces with the vacuum at $z=0$ and $L$. The sheet current density in this planar conductor $\vec{J}$ is related to the electric field $\vec{E}$ via ohm's law

\begin{equation}
\vec{J}=\mathbf{\sigma}\vec{E},
\end{equation}

\noindent where the conductivity tensor is given by

\begin{equation}
\mathbf{\sigma}=
 \left[ \begin{array}{cc}
\sigma_{xx} & -\sigma_{xy} \\
\sigma_{xy} & \sigma_{yy} 
\end{array} \right]=\frac{\sigma_0}{1+B_z^2\mu^2}
 \left[ \begin{array}{cc}
1 & -B_z\mu \\
B_z\mu & 1
\end{array} \right].
\label{CONDEQ}
\end{equation}

\noindent Here, $\sigma_0$ is the zero-field conductivity, $B_z$ is perpendicular magnetic field and $\mu$ is planar mobility. These expressions, resulting from the electron equations of motion including Coulomb and Lorentz forces, are equally valid for massive or massless Dirac electrons, despite the topological protection against backscattering in a TI surface state. We have ignored the effects of a Zeeman gap at the spin-degenerate Dirac point on $\mu$, so our results should be interpreted carefully in the pathological case where this energy scale is dominant (at low temperatures in high magnetic fields when the Fermi energy is at the Dirac point). 

\begin{figure}
\centering
\includegraphics[width=2.3in, height=2in]{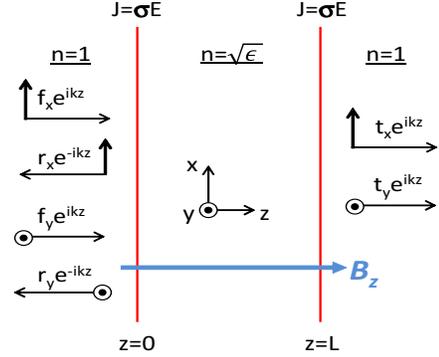}
\caption{Geometry of the normally incident ($f_{x,y}$), reflected ($r_{x,y}$), and transmitted ($t_{x,y}$) electric fields of propagating plane waves connected by scattering from two coherently coupled conductive surfaces on an infinite dielectric slab in a perpendicular magnetic field $B_z$.
\label{FIGSLAB}}
\end{figure}

As illustrated in Fig. \ref{FIGSLAB}, a plane wave normally-incident from the left with linear polarization components $f_x$ and $f_y$ leads to a reflected wave with complex amplitudes $r_x$ and $r_y$, and a transmitted wave on the other side of the slab with complex amplitudes $t_x$ and $t_y$. In a basis defined by \textit{polarization axis} $\otimes$ \textit{propagation direction}, we can describe a connection between the electric fields on the left side of the resonator with those on the right as  

\begin{equation}
 \left[ \begin{array}{c}
f_x \\
r_x \\
f_y \\
r_y  \end{array} \right] = \mathbf{\bar{M}}
 \left[ \begin{array}{c}
t_x \\
0 \\
t_y \\
0  \end{array} \right],
\end{equation}

\noindent where $\mathbf{\bar{M}}$ is a $4\times 4$ transfer matrix determined by the Maxwell boundary conditions at $z=0$ and $L$ and coherent propagation through the bulk. In a basis that separates the known quantities (incoming forward-propagating $f_x$ and $f_y$) from the four unknowns ($r_x$, $t_x$, $r_y$, and $t_y$), we have the equivalent linear system

\begin{equation}
 \left[ \begin{array}{cccc}
0 & \bar{M}_{11} & 0 & \bar{M}_{13} \\
-1 & \bar{M}_{21} & 0 & \bar{M}_{23} \\
0 & \bar{M}_{31} & 0 & \bar{M}_{33} \\
0 & \bar{M}_{41} & -1 & \bar{M}_{43} 
\end{array} \right]
 \left[ \begin{array}{c}
r_x \\
t_x \\
r_y \\
t_y  \end{array} \right] = 
 \left[ \begin{array}{c}
f_x \\
0 \\
f_y \\
0  \end{array} \right],
\label{TMATEQN}
\end{equation}

\noindent where $\bar{M}_{ij}$ is the element of the transfer matrix $\mathbf{\bar{M}}$ in the $i^{th}$ row and $j^{th}$ column. By inverting the matrix in Eq. (\ref{TMATEQN}), we can calculate the observed transmittance and reflectance from the intensity of waves of both polarizations propagating into the vacuum, $T_{x,y}=|t_{x,y}|^2$ and $R_{x,y}=|r_{x,y}|^2$, respectively.

\begin{figure}%[h!]
\centering
\includegraphics[width=3.4in]{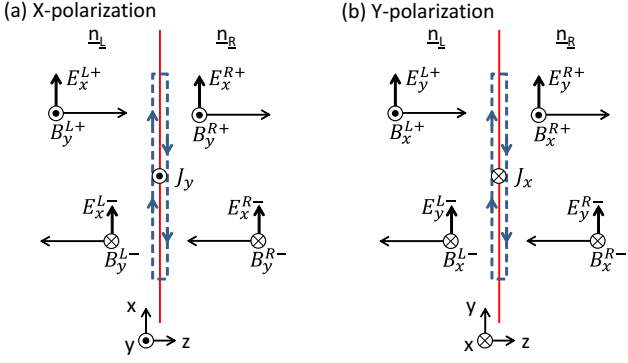}
\caption{(a) Geometry of x-polarized E-field plane waves associated with forward (superscript $+$) and backward (superscript $-$) propagating waves on the left (superscript $L$) and right (superscript $R$) of a conductive boundary. (b) Geometry for y--polarized E-field plane waves. The dashed box of negligible widths intersecting the boundary indicate integration paths used to obtain the boundary conditions.
\label{FIGBC}}
\end{figure}

The appropriate boundary conditions on electric fields in the three regions (left, inside, and right of the slab) are provided by the integral versions of Faraday's and Ampere's laws applied to a closed path intersecting a general boundary and with vanishing enclosed field fluxes (but nonzero enclosed current), as illustrated in Fig. \ref{FIGBC}. For the wave with electric field polarization along $x$ [see Fig. \ref{FIGBC}(a)], we have

\begin{align}
\label{XFARADAYEQ}
E_x^{L+}+E_x^{L-}&=E_x^{R+}+E_x^{R-} \text{ [Faraday], and} \\
B_y^{L+}-B_y^{L-}&=B_y^{R+}-B_y^{R-}+\mu_0J_x \text{ [Ampere],} \nonumber
\end{align}

\noindent where $\mu_0$ is the vacuum magnetic permeability, and the superscript indicates the field amplitude on the left ($L$) or right ($R$) side of the interface, and refers to either the forward ($+$) or backward ($-$) propagating plane wave. The complex phase $e^{\pm ikz}$ where $+(-)$ is for forward (backward) traveling waves has been suppressed. Note the sign change on the backward-propagating wave to account for the correct direction of power flow: $\vec{k}\parallel \vec{E}\times\vec{B}$. Faraday's law applied to plane waves also gives us the relationship between the amplitudes of the perpendicular $E$ and $B$ components as $B=\frac{n}{c}E$, allowing us to write Ampere's boundary condition above as

\begin{align}
\frac{n_L}{c}(E_x^{L+}-E_x^{L-})&=\nonumber\\
\frac{n_R}{c}(E_x^{R+}-&E_x^{R-})+\mu_0(\sigma_{xx}E_x-\sigma_{xy}E_y),
\label{XAMPEREEQ}
\end{align}

\noindent where $n_{L(R)}$ is the index of refraction on the left(right) of the interface and we have inserted the appropriate component of Ohm's law using Eq. (\ref{CONDEQ}). To preserve left-right symmetry, we will choose $E_x=\frac{1}{2}(E_x^{L+}+E_x^{L-})+\frac{1}{2}(E_x^{R+}+E_x^{R-})$, as allowed by Faraday's law boundary condition, Eq. (\ref{XFARADAYEQ}). Likewise, we also take $E_y=\frac{1}{2}(E_y^{L+}+E_y^{L-})+\frac{1}{2}(E_y^{R+}+E_y^{R-})$.

The $y$-polarized wave similarly has boundary conditions on the fields shown in Fig. \ref{FIGBC}(b) as

\begin{equation}
\label{YFARADAYEQ}
E_y^{L+}+E_y^{L-}=E_y^{R+}+E_y^{R-} \text{ [Faraday], and}
\end{equation}

\vspace{-20pt}

\begin{align}
\frac{n_L}{c}(-E_y^{L+}+E_y^{L-})=\frac{n_R}{c}&(-E_y^{R+}+E_y^{R-})-\nonumber \\
&\mu_0(\sigma_{yy}E_y+\sigma_{xy}E_x) \text{ [Ampere]}.
\label{YAMPEREEQ}
\end{align}

We therefore have four coupled linear equations (\ref{XFARADAYEQ}, \ref{XAMPEREEQ}, \ref{YFARADAYEQ}, and \ref{YAMPEREEQ}) that can be compactly expressed by separating the left ($L$) and right ($R$) fields as

\begin{equation}
\mathbf{M_L}\left[ \begin{array}{c}
E^{L+}_x \\
E^{L-}_x \\
E^{L+}_y \\
E^{L-}_y  \end{array} \right] = 
\mathbf{M_R}\left[ \begin{array}{c}
E^{R+}_x \\
E^{R-}_x \\
E^{R+}_y \\
E^{R-}_y  \end{array} \right] 
\label{EQ_leftright}
\end{equation}

\noindent with matrices

\begin{widetext}

\begin{equation}
\mathbf{M_{L(R)}}=
\left[ \begin{array}{cccc}
e^{ik_{L(R)}z} & e^{-ik_{L(R)}z} & 0 & 0 \\*
(\frac{n_{L(R)}}{c}\mp\frac{\mu_0}{2}\sigma_{xx})e^{ik_{L(R)}z} & (-\frac{n_{L(R)}}{c}\mp\frac{\mu_0}{2}\sigma_{xx})e^{-ik_{L(R)}z} & \pm\frac{\mu_0}{2}\sigma_{xy}e^{ik_{L(R)}z} & \pm\frac{\mu_0}{2}\sigma_{xy}e^{-ik_{L(R)}z}\\*
0 & 0 & e^{ik_{L(R)}z} & e^{-ik_{L(R)}z} \\*
\pm\frac{\mu_0}{2}\sigma_{xy}e^{ik_{L(R)}z} & \pm\frac{\mu_0}{2}\sigma_{xy}e^{-ik_{L(R)}z} & (-\frac{n_{L(R)}}{c}\pm\frac{\mu_0}{2}\sigma_{yy})e^{ik_{L(R)}z} & (\frac{n_{L(R)}}{c}\pm\frac{\mu_0}{2}\sigma_{yy})e^{-ik_{L(R)}z} 
\end{array} \right],
\label{EQ_MR}
\end{equation}

\showMLMR{
\begin{equation}
\mathbf{ M_L}=
\left[ \begin{array}{cccc}
e^{ik_Lz} & e^{-ik_Lz} & 0 & 0 \\*
(\frac{n_L}{c}-\frac{\mu_0}{2}\sigma_{xx})e^{ik_Lz} & (-\frac{n_L}{c}-\frac{\mu_0}{2}\sigma_{xx})e^{-ik_Lz} & \frac{\mu_0}{2}\sigma_{xy}e^{ik_Lz} & \frac{\mu_0}{2}\sigma_{xy}e^{-ik_Lz}\\*
0 & 0 & e^{ik_Lz} & e^{-ik_Lz} \\*
\frac{\mu_0}{2}\sigma_{xy}e^{ik_Lz} & \frac{\mu_0}{2}\sigma_{xy}e^{-ik_Lz} & (-\frac{n_L}{c}+\frac{\mu_0}{2}\sigma_{yy})e^{ik_Lz} & (\frac{n_L}{c}+\frac{\mu_0}{2}\sigma_{yy})e^{-ik_Lz} 
\end{array} \right]
%\label{EQ_MR}
\end{equation}
\noindent and
\begin{equation}
\mathbf{M_R}=
\left[ \begin{array}{cccc}
e^{ik_Rz} & e^{-ik_Rz} & 0 & 0 \\*
(\frac{n_R}{c}+\frac{\mu_0}{2}\sigma_{xx})e^{ik_Rz} & (-\frac{n_R}{c}+\frac{\mu_0}{2}\sigma_{xx})e^{-ik_Rz} & -\frac{\mu_0}{2}\sigma_{xy}e^{ik_Rz} & -\frac{\mu_0}{2}\sigma_{xy}e^{-ik_Rz}\\*
0 & 0 & e^{ik_Rz} & e^{-ik_Rz} \\*
-\frac{\mu_0}{2}\sigma_{xy}e^{ik_Rz} & -\frac{\mu_0}{2}\sigma_{xy}e^{-ik_Rz} & (-\frac{n_R}{c}-\frac{\mu_0}{2}\sigma_{yy})e^{ik_Rz} & (\frac{n_R}{c}-\frac{\mu_0}{2}\sigma_{yy})e^{-ik_Rz} 
\end{array} \right].
\end{equation}
}

\end{widetext}

\noindent where $\pm$ and $\mp$ correspond to $\mathbf{M_L}$ (sign above) and $\mathbf{M_R}$ (sign below).
Note that the exponential phase factors $e^{ikz}$ have now been included. Here, $k_L=n_Lk_0$, $k_R=n_Rk_0$, where $k_0=\omega/c$ is the free-space wavenumber.

From these definitions, we can construct the transfer matrix $\mathbf{M}^{(i)}=\mathbf{M_L}^{-1}\mathbf{M_R}$ for the $i=1,2$ two interfaces at $z=0$ and $z=L$, with appropriate substitutions as shown in Table \ref{TABLE}. The total transfer matrix of this slab dielectric is then merely the matrix product of the two interfaces in appropriate order: $\mathbf{\bar{M}}=\mathbf{M^{(1)}}\mathbf{M^{(2)}}$, to be used in Eq. (\ref{TMATEQN}) to obtain the amplitudes $t_x$, $t_y$, $r_x$, and $r_y$ for a chosen wavenumber $k$.

\begin{table}
\caption{Parameters for interface transfer matrices}
\begin{tabular}{c| ccc}
\hline
$i$ & $z$ & $n_L$ & $n_R$ \\
\hline\hline
1 & 0 & 1 & $\sqrt{\epsilon}$ \\
2 & $L$ & $\sqrt{\epsilon}$ & 1 \\
\hline
\end{tabular}
\label{TABLE}
\end{table}

Since we have a nonzero off-diagonal conductivity $\sigma_{xy}=\frac{\sigma_0B_z\mu}{1+B_z^2\mu^2}$ and modify the diagonal conductivity $\sigma_{xx}=\sigma_{yy}=\frac{\sigma_0}{1+B_z^2\mu^2}$ by controlling a static magnetic field $B_z$, we must also consider polarization rotation by a bulk Faraday effect. We can include this nonreciprocal mechanism by inserting 

\begin{equation}
\mathbf{M_F}=
\left[ \begin{array}{c c c c}
\cos\beta & 0 & \sin\beta & 0\\
0 & \cos\beta  & 0 & -\sin\beta \\
 -\sin\beta & 0 & \cos\beta & 0 \\
 0 & \sin\beta & 0 & \cos\beta 
\end{array} \right]
\end{equation}

\noindent between $\mathbf{M^{(1)}}$ and $\mathbf{M^{(2)}}$, where the Faraday angle $\beta=\mathcal{V}B_zL$. The Verdet constant $\mathcal{V}$ has typical values as high as $1$~rad/Tcm in the optical regime, but far lower for the much longer microwave wavelengths to be used in the present application.

\emph{Results--} First, we examine the results of simulating transmission and reflection spectra as a function of magnetic field $B_z$ in the absence of a conducting surface state, but with nonzero Verdet constant. We choose a relative permittivity $\epsilon=600$ (appropriate for the Kondo topological insulator SmB$_6$\cite{Dzero_PRL2010, Kim_SciRep2013, Thomas_arxiv2013, Yee_arxiv2013, Zhu_PRL2013, Neupane_arxiv2013} at low temperature; see Refs. \onlinecite{Dressel_PhysicaB1999} and \onlinecite{Gorshunov_PRB1999}). All simulations use a linearly-polarized incident field $f_x=1$ and $f_y=0$. 

\begin{figure}
\centering
\includegraphics[height=4.5in]{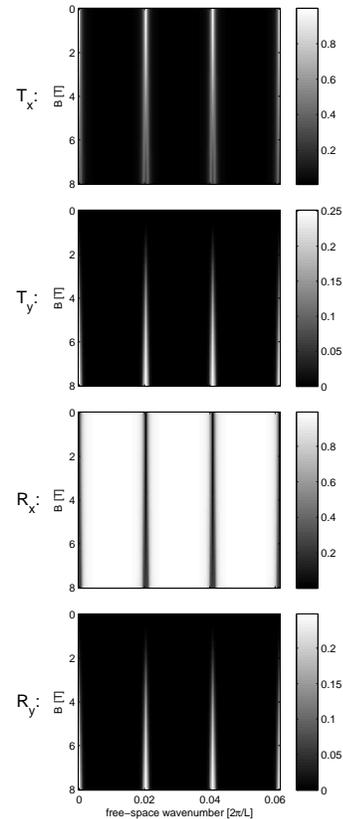}
\caption{Calculation of transmission and reflection coefficient for both $x$ and $y$ linear polarizations including Faraday effect in the $L=1$mm-long bulk, using Verdet constant $V=10^{-1}$~rad/Tcm, with an incident plane wave polarized along $x$. Here, there is no surface state. The index of refraction between conductive layers is determined by the relative permittivity $\epsilon=600$. Faraday effect merely mixes the polarization states but does not lead to any qualitatively different $T_y$ spectra.
\label{FIGTR1}}
\end{figure}

As shown in Fig. \ref{FIGTR1}, dielectric resonances\cite{Richtmyer_JAP1939} at free-space wavenumbers $m\frac{\pi}{\sqrt{\epsilon}L}$, where $m=1,2,...$ are clearly evident in colinear transmittance $T_x$. The nonreciprocal polarization rotation from bulk Faraday effect (using $\mathcal{V}=10^{-1}$~rad/Tcm through the $L=1$~mm dielectric resonator) induced by increasing $B_z$ mixes power into the orthogonal direction ($T_y$). Importantly, however, these maxima occur at the same wavenumbers as the dielectric resonances in $T_x$ and are qualitatively identical.

\begin{figure}
\centering
\includegraphics[width=3.5in, height=5.5in]{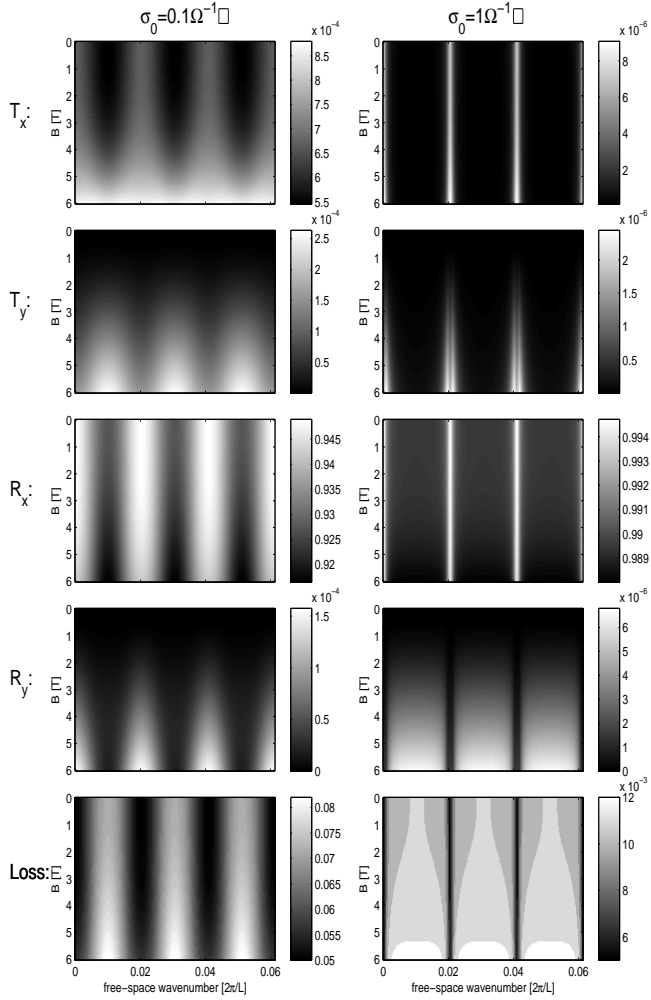}
\caption{Calculation of transmission and reflection coefficient for both $x$ and $y$ linear polarizations in the absence of bulk Faraday effect in the $L=1$mm resonator, with an incident plane wave polarized along $x$. Here, the index of refraction between conductive layers is determined by the relative permittivity $\epsilon=600$. The unique signature of surface conduction (a suppression at resonance in a magnetic field such that $B_z\mu\approx 1$, where $\mu$ is mobility) appears as a maxima at anti-resonance wavevectors for low conductivity on the left, or a central suppression of metallic resonance peaks for high conductivity on the right) is seen in $T_y$.  
\label{FIGTR0}}
\end{figure}

\begin{figure}
\centering
\includegraphics[width=3in]{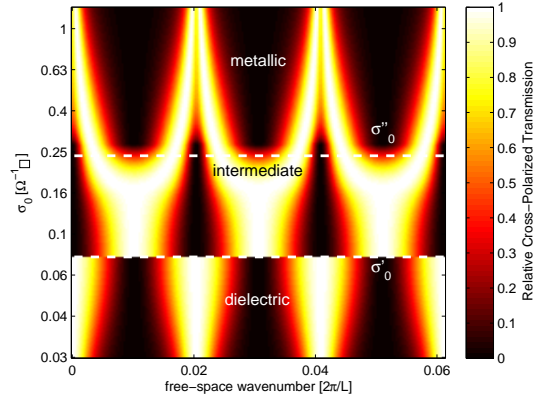}
\caption{Normalized relative transmission spectra of orthogonal linear polarization at magnetic field $B=6$~T vs. surface conductivity $\sigma_0$. At very low conductivity, maxima are due to damped dielectric resonances; at high conductivity maxima are due to metallic resonances with a central suppression. For intermediate conductivity, maxima appear at anti-resonances due to strong electric fields at the resonator boundary inducing orthogonal currents. Transition conductivities between the three regimes are indicated by horizontal dashed lines.
\label{FIGspectra}}
\end{figure}

When the conducting surface state is included, we obtain example results as shown in Fig. \ref{FIGTR0}. Here, $\mu=10^3$~cm$^2$/Vs (so that $B_z\mu=1$ at 10T) and $\mathcal{V}=0$. For substantial $\sigma_{xx}$ but negligible $\sigma_{xy}$ (upper part of plots in right column), the transmission maxima in $T_x$ are no longer dielectric resonances but rather due to the metallic boundary conditions, resulting in higher reflectance $R_x$.

In a magnetic field such that $B_z\mu$ approaches unity, $\sigma_{xy}$ becomes comparable to $\sigma_{xx}$. For high surface conductivity $\sigma_0=1\Omega^{-1}\Box$ (right column), the resonances in $T_x$ are then coupled with a central split into $T_y$; the splitting can either be enhanced or suppressed by bulk Faraday effect, depending on the sign of a nonzero Verdet constant (not shown). This splitting can be understood by considering that for perfectly conducting boundary conditions, the electric field (and hence the perpendicular current proportional to $\sigma_{xy}$) at the resonator surface goes to zero. In the $\sigma_0\rightarrow\infty$ limit, all field coupling to $E_y$ then vanishes.  

For intermediate surface conductivity, comparable to experimental observations in SmB$_6$ ($\sigma_0=0.1\Omega^{-1}\Box$, left plots in Fig. \ref{FIGTR0}), the magnetic field also couples the transmitted electric field into the $y$ orientation, but a significant difference is evident: maxima at \textbf{anti-resonances} $k=(m-\frac{1}{2})\frac{\pi}{\sqrt{\epsilon}L}$ can be seen in $T_y$. This phenomena is explained by the fact that the electric field and resulting induced surface current for these wavenumbers is \emph{maximized} at the boundary. When the overall conductance is not high enough to completely suppress these modes, substantial polarization mixing will result in this qualitatively unique perpendicularly polarized spectra that occurs in an intermediate regime between dielectric and metallic resonance. 

\emph{Discussion-} To highlight the role of conductivity in determining the spectral behavior of the cross-polarized channel, we plot the relative transmission $0\leq\frac{T_y-\min{T_y}}{\max{T_y}-\min{T_y}}\leq 1$ for a fixed $B_z=$6T in Fig. \ref{FIGspectra}. Because the wavenumbers for maxima and minima are abruptly exchanged at the boundary between fully dielectric resonance ($\sigma_0\approx 0$) and the intermediate regime ($\sigma_0\approx 0.1\Omega^{-1}\Box$), there must be a conductivity $\sigma_0'$ for which the transmission spectrum is frequency-independent. This occurs when the conductive interface serves to impedance match between dielectric and vacuum, eliminating reflection and fully suppressing all interference from multiple paths. By solving Eqns. \ref{EQ_leftright}-\ref{EQ_MR} for a single interface with $n_L=\sqrt{\epsilon}$, $n_R=1$ and $E_{x,y}^{L-,R-}=0$, we can analytically determine the value where this occurs at

\begin{equation}
\sigma_0'=\frac{\sqrt{\epsilon}-1}{c\mu_0}(1+B_z^2\mu^2).
\label{EQ_sigprime}
\end{equation}

\noindent For $\epsilon=600$ here, $\sigma_0'\approx 0.085\Omega^{-1}\Box$, matching the results (highlighted with a horizontal dashed line) in Fig. \ref{FIGspectra}.  

The transition conductivity between intermediate and metallic regimes at $\approx 0.25\Omega^{-1}\Box$ in Fig. \ref{FIGspectra} is less abrupt. However, we can still define a characteristic value at a point of low enough conductivity $\sigma_0^{{'}{'}}$ where the metallic resonances broaden to a width comparable to the mode spacing $\Delta k= \frac{\pi}{\sqrt{\epsilon}L}$ so that spectral weight can accumulate at the nodes. This occurs when the cavity Q-factor

\begin{equation}
Q=\frac{\omega U}{P}\approx\frac{k_0c(\epsilon\epsilon_0E^2+\frac{1}{\mu_0c^2}E^2)L}{\sigma_{xx} E^2}\approx\frac{k_0}{\Delta k},
\end{equation}

\noindent where $U$ is the average stored energy in the dielectric (from two counter-propagating planewaves) and $P$ is the power loss due to Joule heating at the conductive boundaries. Our condition gives 

\begin{equation}
\sigma_0^{{'}{'}}\approx\frac{\pi(\epsilon+1)}{\sqrt{\epsilon}\mu_0c}(1+B_z^2\mu^2),
\label{EQ_sigdprimea}
\end{equation}

\noindent which has a value of $\approx 0.28\Omega^{-1}\Box$ for the parameters used in Fig. \ref{FIGspectra}, again in apparently good agreement with the calculated results. It is worth noting that although $\sigma_0'\rightarrow 0$ for $\epsilon\rightarrow 1$, $\sigma_0^{{'}{'}}$ remains finite as expected for a lossy metal cavity. While this expression is satisfactory in predicting the transition conductivity for the high dielectric constant here, in the low-$\epsilon$ limit a better approximation is rather given by the onset of metallic resonance, requiring a substantial reflected field relative to the incident field. Again using Eqns. \ref{EQ_leftright}-\ref{EQ_MR}, but now with $|E_x^{L-}|=\frac{1}{2}E_x^{L+}$, we find

\begin{equation}
\sigma_0^{{'}{'}}\approx\frac{3\sqrt{\epsilon}-1}{\mu_0c}(1+B_z^2\mu^2).
\label{EQ_sigdprimeb}
\end{equation}

\noindent Note that the asymptotic behavior of Eqs. \ref{EQ_sigdprimea} and \ref{EQ_sigdprimeb} in $\epsilon\rightarrow\infty$ is nearly identical.

\begin{figure}
\centering
\includegraphics[width=3in]{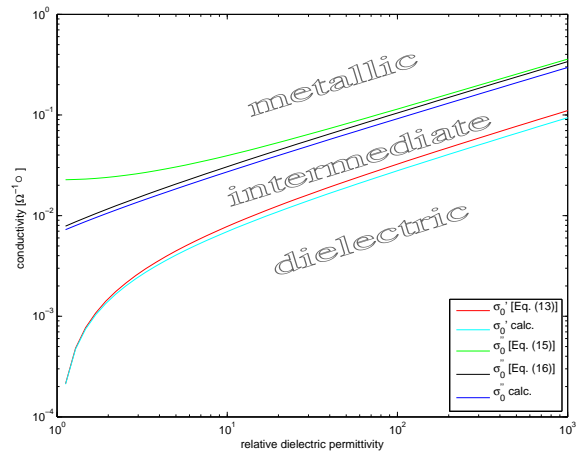}
\caption{Phase diagram at $B_z=6$T identifying the conditions necessary for the three regimes of cross-polarized transmission spectrum characteristics. Calculated boundaries are compared with the analytic expressions in Eqs. \ref{EQ_sigprime}, \ref{EQ_sigdprimea}, and \ref{EQ_sigdprimeb} derived from properties of the single interface, Eq. \ref{EQ_leftright}.  
\label{FIGeps}}
\end{figure}

Figure \ref{FIGeps} compares the analytic boundaries defined in Eqs. \ref{EQ_sigprime}, \ref{EQ_sigdprimea}, and \ref{EQ_sigdprimeb} to boundaries calculated from a binomial search for transitions in relative cross-polarized transmission data (generated using the transfer matrix approach and similar to Fig. \ref{FIGspectra}) as a function of dielectric permittivity $\epsilon$. In this phase diagram, the dielectric, intermediate, and metallic regimes are clearly defined, and the limitation of Eq. \ref{EQ_sigdprimea} at low-$\epsilon$ is evident.   

\emph{Conclusion-} We conclude that the orthogonal linearly polarized transmission spectra in a magnetic field carries an unambiguous signature of a sufficiently conductive metallic surface state: suppression of transmission at resonance leading either to a split peak for highly conductive surfaces, or a maxima at anti-resonance for intermediate conductivity.

These 1-dimensional simulations are most applicable to experiments on planar samples and thin films, thick enough that the optical pathlength is at least half the wavelength of a photon with less than the bulk bandgap energy (to prohibit interband excitation), but thin and smooth enough to maintain resonance peak separation and suppress broadening. 

Incorporating more elaborate modifications to the present model should be straightforward. For instance, simulating the regime where quantized Hall conductivity\cite{Maciejko_PRL2010, Tse_PRL2010, Tse_PRB2011} is expected can be treated with a suitable adjustment to the conductivity tensor, Eq. \ref{CONDEQ}.

\begin{acknowledgments}
We acknowledge helpful discussions with S. Anlage, J. Paglione, J.D. Sau, H.D. Drew and D. Weile.
\end{acknowledgments}

%\bibliography{TIresonator_v5}

\begin{thebibliography}{22}%
\makeatletter
\providecommand \@ifxundefined [1]{%
 \@ifx{#1\undefined}
}%
\providecommand \@ifnum [1]{%
 \ifnum #1\expandafter \@firstoftwo
 \else \expandafter \@secondoftwo
 \fi
}%
\providecommand \@ifx [1]{%
 \ifx #1\expandafter \@firstoftwo
 \else \expandafter \@secondoftwo
 \fi
}%
\providecommand \natexlab [1]{#1}%
\providecommand \enquote  [1]{``#1''}%
\providecommand \bibnamefont  [1]{#1}%
\providecommand \bibfnamefont [1]{#1}%
\providecommand \citenamefont [1]{#1}%
\providecommand \href@noop [0]{\@secondoftwo}%
\providecommand \href [0]{\begingroup \@sanitize@url \@href}%
\providecommand \@href[1]{\@@startlink{#1}\@@href}%
\providecommand \@@href[1]{\endgroup#1\@@endlink}%
\providecommand \@sanitize@url [0]{\catcode `\\12\catcode `\$12\catcode
  `\&12\catcode `\#12\catcode `\^12\catcode `\_12\catcode `\%12\relax}%
\providecommand \@@startlink[1]{}%
\providecommand \@@endlink[0]{}%
\providecommand \url  [0]{\begingroup\@sanitize@url \@url }%
\providecommand \@url [1]{\endgroup\@href {#1}{\urlprefix }}%
\providecommand \urlprefix  [0]{URL }%
\providecommand \Eprint [0]{\href }%
\providecommand \doibase [0]{http://dx.doi.org/}%
\providecommand \selectlanguage [0]{\@gobble}%
\providecommand \bibinfo  [0]{\@secondoftwo}%
\providecommand \bibfield  [0]{\@secondoftwo}%
\providecommand \translation [1]{[#1]}%
\providecommand \BibitemOpen [0]{}%
\providecommand \bibitemStop [0]{}%
\providecommand \bibitemNoStop [0]{.\EOS\space}%
\providecommand \EOS [0]{\spacefactor3000\relax}%
\providecommand \BibitemShut  [1]{\csname bibitem#1\endcsname}%
\let\auto@bib@innerbib\@empty
%</preamble>
\bibitem [{\citenamefont {Hasan}\ and\ \citenamefont
  {Kane}(2010)}]{Hasan_RMP2010}%
  \BibitemOpen
  \bibfield  {author} {\bibinfo {author} {\bibfnamefont {M.~Z.}\ \bibnamefont
  {Hasan}}\ and\ \bibinfo {author} {\bibfnamefont {C.~L.}\ \bibnamefont
  {Kane}},\ }\href {\doibase 10.1103/RevModPhys.82.3045} {\bibfield  {journal}
  {\bibinfo  {journal} {Rev. Mod. Phys.}\ }\textbf {\bibinfo {volume} {82}},\
  \bibinfo {pages} {3045} (\bibinfo {year} {2010})}\BibitemShut {NoStop}%
\bibitem [{\citenamefont {Portis}\ and\ \citenamefont
  {Teaney}(1958)}]{Portis_JAP1958}%
  \BibitemOpen
  \bibfield  {author} {\bibinfo {author} {\bibfnamefont {A.~M.}\ \bibnamefont
  {Portis}}\ and\ \bibinfo {author} {\bibfnamefont {D.}~\bibnamefont
  {Teaney}},\ }\href {\doibase http://dx.doi.org/10.1063/1.1723027} {\bibfield
  {journal} {\bibinfo  {journal} {J. Appl. Phys.}\ }\textbf {\bibinfo {volume}
  {29}},\ \bibinfo {pages} {1692} (\bibinfo {year} {1958})}\BibitemShut
  {NoStop}%
\bibitem [{\citenamefont {Moore}(1973)}]{Moore_RSI1973}%
  \BibitemOpen
  \bibfield  {author} {\bibinfo {author} {\bibfnamefont {W.~S.}\ \bibnamefont
  {Moore}},\ }\href {\doibase http://dx.doi.org/10.1063/1.1686072} {\bibfield
  {journal} {\bibinfo  {journal} {Rev. Sci. Instr.}\ }\textbf {\bibinfo
  {volume} {44}},\ \bibinfo {pages} {158} (\bibinfo {year} {1973})}\BibitemShut
  {NoStop}%
\bibitem [{\citenamefont {Suhl}\ and\ \citenamefont
  {Walker}(1952)}]{Suhl_PR1952}%
  \BibitemOpen
  \bibfield  {author} {\bibinfo {author} {\bibfnamefont {H.}~\bibnamefont
  {Suhl}}\ and\ \bibinfo {author} {\bibfnamefont {L.~R.}\ \bibnamefont
  {Walker}},\ }\href {\doibase 10.1103/PhysRev.86.122} {\bibfield  {journal}
  {\bibinfo  {journal} {Phys. Rev.}\ }\textbf {\bibinfo {volume} {86}},\
  \bibinfo {pages} {122} (\bibinfo {year} {1952})}\BibitemShut {NoStop}%
\bibitem [{Tru()}]{Trukhan_REEP1966}%
  \BibitemOpen
  \href@noop {} {\ }\BibitemShut {NoStop}%
\bibitem [{\citenamefont {Fletcher}(1976)}]{Fletcher_JPhysE1976}%
  \BibitemOpen
  \bibfield  {author} {\bibinfo {author} {\bibfnamefont {J.~R.}\ \bibnamefont
  {Fletcher}},\ }\href {http://stacks.iop.org/0022-3735/9/i=6/a=017} {\bibfield
   {journal} {\bibinfo  {journal} {J. Phys. E: Sci. Instr.}\ }\textbf {\bibinfo
  {volume} {9}},\ \bibinfo {pages} {481} (\bibinfo {year} {1976})}\BibitemShut
  {NoStop}%
\bibitem [{\citenamefont {Eley}\ and\ \citenamefont
  {Lockhart}(1983)}]{Eley_JPhysE1983}%
  \BibitemOpen
  \bibfield  {author} {\bibinfo {author} {\bibfnamefont {D.~D.}\ \bibnamefont
  {Eley}}\ and\ \bibinfo {author} {\bibfnamefont {N.~C.}\ \bibnamefont
  {Lockhart}},\ }\href {http://stacks.iop.org/0022-3735/16/i=1/a=010}
  {\bibfield  {journal} {\bibinfo  {journal} {J. Phys. E: Sci. Instr.}\
  }\textbf {\bibinfo {volume} {16}},\ \bibinfo {pages} {47} (\bibinfo {year}
  {1983})}\BibitemShut {NoStop}%
\bibitem [{\citenamefont {Aguilar}\ \emph {et~al.}(2011)\citenamefont
  {Aguilar}, \citenamefont {Stier}, \citenamefont {Liu}, \citenamefont
  {Bilbro}, \citenamefont {George}, \citenamefont {Bansal}, \citenamefont {Wu},
  \citenamefont {Cerne}, \citenamefont {Markelz}, \citenamefont {Oh},\ and\
  \citenamefont {Armitage}}]{Aguilar_PRL2011}%
  \BibitemOpen
  \bibfield  {author} {\bibinfo {author} {\bibfnamefont {R.~V.}\ \bibnamefont
  {Aguilar}}, \bibinfo {author} {\bibfnamefont {A.~V.}\ \bibnamefont {Stier}},
  \bibinfo {author} {\bibfnamefont {W.}~\bibnamefont {Liu}}, \bibinfo {author}
  {\bibfnamefont {L.~S.}\ \bibnamefont {Bilbro}}, \bibinfo {author}
  {\bibfnamefont {D.~K.}\ \bibnamefont {George}}, \bibinfo {author}
  {\bibfnamefont {N.}~\bibnamefont {Bansal}}, \bibinfo {author} {\bibfnamefont
  {L.}~\bibnamefont {Wu}}, \bibinfo {author} {\bibfnamefont {J.}~\bibnamefont
  {Cerne}}, \bibinfo {author} {\bibfnamefont {A.~G.}\ \bibnamefont {Markelz}},
  \bibinfo {author} {\bibfnamefont {S.}~\bibnamefont {Oh}}, \ and\ \bibinfo
  {author} {\bibfnamefont {N.~P.}\ \bibnamefont {Armitage}},\ }\href
  {http://dx.doi.org/10.1103/PhysRevLett.108.087403} {\bibfield  {journal}
  {\bibinfo  {journal} {Phys. Rev. Lett.}\ }\textbf {\bibinfo {volume} {108}},\
  \bibinfo {pages} {087403} (\bibinfo {year} {2011})}\BibitemShut {NoStop}%
\bibitem [{\citenamefont {Maciejko}\ \emph {et~al.}(2010)\citenamefont
  {Maciejko}, \citenamefont {Qi}, \citenamefont {Drew},\ and\ \citenamefont
  {Zhang}}]{Maciejko_PRL2010}%
  \BibitemOpen
  \bibfield  {author} {\bibinfo {author} {\bibfnamefont {J.}~\bibnamefont
  {Maciejko}}, \bibinfo {author} {\bibfnamefont {X.-L.}\ \bibnamefont {Qi}},
  \bibinfo {author} {\bibfnamefont {H.~D.}\ \bibnamefont {Drew}}, \ and\
  \bibinfo {author} {\bibfnamefont {S.-C.}\ \bibnamefont {Zhang}},\ }\href
  {\doibase 10.1103/PhysRevLett.105.166803} {\bibfield  {journal} {\bibinfo
  {journal} {Phys. Rev. Lett.}\ }\textbf {\bibinfo {volume} {105}},\ \bibinfo
  {pages} {166803} (\bibinfo {year} {2010})}\BibitemShut {NoStop}%
\bibitem [{\citenamefont {Shuvaev}\ \emph {et~al.}(2011)\citenamefont
  {Shuvaev}, \citenamefont {Astakhov}, \citenamefont {Pimenov}, \citenamefont
  {Br\"une}, \citenamefont {Buhmann},\ and\ \citenamefont
  {Molenkamp}}]{Shuvaev_PRL2011}%
  \BibitemOpen
  \bibfield  {author} {\bibinfo {author} {\bibfnamefont {A.~M.}\ \bibnamefont
  {Shuvaev}}, \bibinfo {author} {\bibfnamefont {G.~V.}\ \bibnamefont
  {Astakhov}}, \bibinfo {author} {\bibfnamefont {A.}~\bibnamefont {Pimenov}},
  \bibinfo {author} {\bibfnamefont {C.}~\bibnamefont {Br\"une}}, \bibinfo
  {author} {\bibfnamefont {H.}~\bibnamefont {Buhmann}}, \ and\ \bibinfo
  {author} {\bibfnamefont {L.~W.}\ \bibnamefont {Molenkamp}},\ }\href
  {http://dx.doi.org/10.1103/PhysRevLett.106.107404} {\bibfield  {journal}
  {\bibinfo  {journal} {Phys. Rev. Lett.}\ }\textbf {\bibinfo {volume} {106}},\
  \bibinfo {pages} {107404} (\bibinfo {year} {2011})}\BibitemShut {NoStop}%
\bibitem [{\citenamefont {Tkachov}\ and\ \citenamefont
  {Hankiewicz}(2011)}]{Tkachov_PRB2011}%
  \BibitemOpen
  \bibfield  {author} {\bibinfo {author} {\bibfnamefont {G.}~\bibnamefont
  {Tkachov}}\ and\ \bibinfo {author} {\bibfnamefont {E.~M.}\ \bibnamefont
  {Hankiewicz}},\ }\href {\doibase 10.1103/PhysRevB.84.035405} {\bibfield
  {journal} {\bibinfo  {journal} {Phys. Rev. B}\ }\textbf {\bibinfo {volume}
  {84}},\ \bibinfo {pages} {035405} (\bibinfo {year} {2011})}\BibitemShut
  {NoStop}%
\bibitem [{\citenamefont {Tse}\ and\ \citenamefont
  {MacDonald}(2010)}]{Tse_PRL2010}%
  \BibitemOpen
  \bibfield  {author} {\bibinfo {author} {\bibfnamefont {W.-K.}\ \bibnamefont
  {Tse}}\ and\ \bibinfo {author} {\bibfnamefont {A.~H.}\ \bibnamefont
  {MacDonald}},\ }\href {\doibase 10.1103/PhysRevLett.105.057401} {\bibfield
  {journal} {\bibinfo  {journal} {Phys. Rev. Lett.}\ }\textbf {\bibinfo
  {volume} {105}},\ \bibinfo {pages} {057401} (\bibinfo {year}
  {2010})}\BibitemShut {NoStop}%
\bibitem [{\citenamefont {Tse}\ and\ \citenamefont
  {MacDonald}(2011)}]{Tse_PRB2011}%
  \BibitemOpen
  \bibfield  {author} {\bibinfo {author} {\bibfnamefont {W.-K.}\ \bibnamefont
  {Tse}}\ and\ \bibinfo {author} {\bibfnamefont {A.~H.}\ \bibnamefont
  {MacDonald}},\ }\href {\doibase 10.1103/PhysRevB.84.205327} {\bibfield
  {journal} {\bibinfo  {journal} {Phys. Rev. B}\ }\textbf {\bibinfo {volume}
  {84}},\ \bibinfo {pages} {205327} (\bibinfo {year} {2011})}\BibitemShut
  {NoStop}%
\bibitem [{\citenamefont {Dzero}\ \emph {et~al.}(2010)\citenamefont {Dzero},
  \citenamefont {Sun}, \citenamefont {Galitski},\ and\ \citenamefont
  {Coleman}}]{Dzero_PRL2010}%
  \BibitemOpen
  \bibfield  {author} {\bibinfo {author} {\bibfnamefont {M.}~\bibnamefont
  {Dzero}}, \bibinfo {author} {\bibfnamefont {K.}~\bibnamefont {Sun}}, \bibinfo
  {author} {\bibfnamefont {V.}~\bibnamefont {Galitski}}, \ and\ \bibinfo
  {author} {\bibfnamefont {P.}~\bibnamefont {Coleman}},\ }\href {\doibase
  10.1103/PhysRevLett.104.106408} {\bibfield  {journal} {\bibinfo  {journal}
  {Phys. Rev. Lett.}\ }\textbf {\bibinfo {volume} {104}},\ \bibinfo {pages}
  {106408} (\bibinfo {year} {2010})}\BibitemShut {NoStop}%
\bibitem [{\citenamefont {Kim}\ \emph {et~al.}(2013)\citenamefont {Kim},
  \citenamefont {Thomas}, \citenamefont {Grant}, \citenamefont {Botimer},
  \citenamefont {Fisk},\ and\ \citenamefont {Xia}}]{Kim_SciRep2013}%
  \BibitemOpen
  \bibfield  {author} {\bibinfo {author} {\bibfnamefont {D.}~\bibnamefont
  {Kim}}, \bibinfo {author} {\bibfnamefont {S.}~\bibnamefont {Thomas}},
  \bibinfo {author} {\bibfnamefont {T.}~\bibnamefont {Grant}}, \bibinfo
  {author} {\bibfnamefont {J.}~\bibnamefont {Botimer}}, \bibinfo {author}
  {\bibfnamefont {Z.}~\bibnamefont {Fisk}}, \ and\ \bibinfo {author}
  {\bibfnamefont {J.}~\bibnamefont {Xia}},\ }\href
  {http://arxiv.org/abs/1211.6769} {\bibfield  {journal} {\bibinfo  {journal}
  {Scientific Rep.}\ }\textbf {\bibinfo {volume} {3}},\ \bibinfo {pages} {3150}
  (\bibinfo {year} {2013})},\ \Eprint {http://arxiv.org/abs/cond-mat/1211.6769}
  {arXiv:cond-mat/1211.6769} \BibitemShut {NoStop}%
\bibitem [{\citenamefont {Thomas}\ \emph {et~al.}(2013)\citenamefont {Thomas},
  \citenamefont {Kim}, \citenamefont {Chung}, \citenamefont {Grant},
  \citenamefont {Fisk},\ and\ \citenamefont {Xia}}]{Thomas_arxiv2013}%
  \BibitemOpen
  \bibfield  {author} {\bibinfo {author} {\bibfnamefont {S.}~\bibnamefont
  {Thomas}}, \bibinfo {author} {\bibfnamefont {D.}~\bibnamefont {Kim}},
  \bibinfo {author} {\bibfnamefont {S.~B.}\ \bibnamefont {Chung}}, \bibinfo
  {author} {\bibfnamefont {T.}~\bibnamefont {Grant}}, \bibinfo {author}
  {\bibfnamefont {Z.}~\bibnamefont {Fisk}}, \ and\ \bibinfo {author}
  {\bibfnamefont {J.}~\bibnamefont {Xia}},\ }\href
  {http://arxiv.org/abs/1307.4133} {} (\bibinfo {year} {2013}),\ \Eprint
  {http://arxiv.org/abs/cond-mat/1307.4133} {arXiv:cond-mat/1307.4133}
  \BibitemShut {NoStop}%
\bibitem [{\citenamefont {Yee}\ \emph {et~al.}(2013)\citenamefont {Yee},
  \citenamefont {He}, \citenamefont {Soumyanarayanan}, \citenamefont {Kim},
  \citenamefont {Fisk},\ and\ \citenamefont {Hoffman}}]{Yee_arxiv2013}%
  \BibitemOpen
  \bibfield  {author} {\bibinfo {author} {\bibfnamefont {M.~M.}\ \bibnamefont
  {Yee}}, \bibinfo {author} {\bibfnamefont {Y.}~\bibnamefont {He}}, \bibinfo
  {author} {\bibfnamefont {A.}~\bibnamefont {Soumyanarayanan}}, \bibinfo
  {author} {\bibfnamefont {D.-J.}\ \bibnamefont {Kim}}, \bibinfo {author}
  {\bibfnamefont {Z.}~\bibnamefont {Fisk}}, \ and\ \bibinfo {author}
  {\bibfnamefont {J.~E.}\ \bibnamefont {Hoffman}},\ }\href
  {http://arxiv.org/abs/1308.1085} {} (\bibinfo {year} {2013}),\ \Eprint
  {http://arxiv.org/abs/cond-mat/1308.1085} {arXiv:cond-mat/1308.1085}
  \BibitemShut {NoStop}%
\bibitem [{\citenamefont {Zhu}\ \emph {et~al.}(2013)\citenamefont {Zhu},
  \citenamefont {Nicolaou}, \citenamefont {Levy}, \citenamefont {Butch},
  \citenamefont {Syers}, \citenamefont {Wang}, \citenamefont {Paglione},
  \citenamefont {Sawatzky}, \citenamefont {Elfimov},\ and\ \citenamefont
  {Damascelli}}]{Zhu_PRL2013}%
  \BibitemOpen
  \bibfield  {author} {\bibinfo {author} {\bibfnamefont {Z.-H.}\ \bibnamefont
  {Zhu}}, \bibinfo {author} {\bibfnamefont {A.}~\bibnamefont {Nicolaou}},
  \bibinfo {author} {\bibfnamefont {G.}~\bibnamefont {Levy}}, \bibinfo {author}
  {\bibfnamefont {N.~P.}\ \bibnamefont {Butch}}, \bibinfo {author}
  {\bibfnamefont {P.}~\bibnamefont {Syers}}, \bibinfo {author} {\bibfnamefont
  {X.~F.}\ \bibnamefont {Wang}}, \bibinfo {author} {\bibfnamefont
  {J.}~\bibnamefont {Paglione}}, \bibinfo {author} {\bibfnamefont {G.~A.}\
  \bibnamefont {Sawatzky}}, \bibinfo {author} {\bibfnamefont {I.~S.}\
  \bibnamefont {Elfimov}}, \ and\ \bibinfo {author} {\bibfnamefont
  {A.}~\bibnamefont {Damascelli}},\ }\href {\doibase
  10.1103/PhysRevLett.111.216402} {\bibfield  {journal} {\bibinfo  {journal}
  {Phys. Rev. Lett.}\ }\textbf {\bibinfo {volume} {111}},\ \bibinfo {pages}
  {216402} (\bibinfo {year} {2013})}\BibitemShut {NoStop}%
\bibitem [{\citenamefont {Neupane}\ \emph {et~al.}(2013)\citenamefont
  {Neupane}, \citenamefont {Alidoust}, \citenamefont {Xu}, \citenamefont
  {Kondo}, \citenamefont {Kim}, \citenamefont {Liu}, \citenamefont
  {Belopolski}, \citenamefont {Chang}, \citenamefont {Jeng}, \citenamefont
  {Durakiewicz}, \citenamefont {Balicas}, \citenamefont {Lin}, \citenamefont
  {Bansil}, \citenamefont {Shin}, \citenamefont {Fisk},\ and\ \citenamefont
  {Hasan}}]{Neupane_arxiv2013}%
  \BibitemOpen
  \bibfield  {author} {\bibinfo {author} {\bibfnamefont {M.}~\bibnamefont
  {Neupane}}, \bibinfo {author} {\bibfnamefont {N.}~\bibnamefont {Alidoust}},
  \bibinfo {author} {\bibfnamefont {S.-Y.}\ \bibnamefont {Xu}}, \bibinfo
  {author} {\bibfnamefont {T.}~\bibnamefont {Kondo}}, \bibinfo {author}
  {\bibfnamefont {D.-J.}\ \bibnamefont {Kim}}, \bibinfo {author} {\bibfnamefont
  {C.}~\bibnamefont {Liu}}, \bibinfo {author} {\bibfnamefont {I.}~\bibnamefont
  {Belopolski}}, \bibinfo {author} {\bibfnamefont {T.-R.}\ \bibnamefont
  {Chang}}, \bibinfo {author} {\bibfnamefont {H.-T.}\ \bibnamefont {Jeng}},
  \bibinfo {author} {\bibfnamefont {T.}~\bibnamefont {Durakiewicz}}, \bibinfo
  {author} {\bibfnamefont {L.}~\bibnamefont {Balicas}}, \bibinfo {author}
  {\bibfnamefont {H.}~\bibnamefont {Lin}}, \bibinfo {author} {\bibfnamefont
  {A.}~\bibnamefont {Bansil}}, \bibinfo {author} {\bibfnamefont
  {S.}~\bibnamefont {Shin}}, \bibinfo {author} {\bibfnamefont {Z.}~\bibnamefont
  {Fisk}}, \ and\ \bibinfo {author} {\bibfnamefont {M.~Z.}\ \bibnamefont
  {Hasan}},\ }\href {http://arxiv.org/abs/1306.4634} {} (\bibinfo {year}
  {2013}),\ \Eprint {http://arxiv.org/abs/cond-mat/1306.4634}
  {arXiv:cond-mat/1306.4634} \BibitemShut {NoStop}%
\bibitem [{\citenamefont {Dressel}\ \emph {et~al.}(1999)\citenamefont
  {Dressel}, \citenamefont {Gorshunov}, \citenamefont {Sluchanko},
  \citenamefont {Volkov}, \citenamefont {Knebel}, \citenamefont {Loidl},\ and\
  \citenamefont {Kunii}}]{Dressel_PhysicaB1999}%
  \BibitemOpen
  \bibfield  {author} {\bibinfo {author} {\bibfnamefont {M.}~\bibnamefont
  {Dressel}}, \bibinfo {author} {\bibfnamefont {B.}~\bibnamefont {Gorshunov}},
  \bibinfo {author} {\bibfnamefont {N.}~\bibnamefont {Sluchanko}}, \bibinfo
  {author} {\bibfnamefont {A.}~\bibnamefont {Volkov}}, \bibinfo {author}
  {\bibfnamefont {G.}~\bibnamefont {Knebel}}, \bibinfo {author} {\bibfnamefont
  {A.}~\bibnamefont {Loidl}}, \ and\ \bibinfo {author} {\bibfnamefont
  {S.}~\bibnamefont {Kunii}},\ }\href {\doibase
  http://dx.doi.org/10.1016/S0921-4526(98)01485-9} {\bibfield  {journal}
  {\bibinfo  {journal} {Physica B: Condensed Matter}\ }\textbf {\bibinfo
  {volume} {259-261}},\ \bibinfo {pages} {347 } (\bibinfo {year}
  {1999})}\BibitemShut {NoStop}%
\bibitem [{\citenamefont {Gorshunov}\ \emph {et~al.}(1999)\citenamefont
  {Gorshunov}, \citenamefont {Sluchanko}, \citenamefont {Volkov}, \citenamefont
  {Dressel}, \citenamefont {Knebel}, \citenamefont {Loidl},\ and\ \citenamefont
  {Kunii}}]{Gorshunov_PRB1999}%
  \BibitemOpen
  \bibfield  {author} {\bibinfo {author} {\bibfnamefont {B.}~\bibnamefont
  {Gorshunov}}, \bibinfo {author} {\bibfnamefont {N.}~\bibnamefont
  {Sluchanko}}, \bibinfo {author} {\bibfnamefont {A.}~\bibnamefont {Volkov}},
  \bibinfo {author} {\bibfnamefont {M.}~\bibnamefont {Dressel}}, \bibinfo
  {author} {\bibfnamefont {G.}~\bibnamefont {Knebel}}, \bibinfo {author}
  {\bibfnamefont {A.}~\bibnamefont {Loidl}}, \ and\ \bibinfo {author}
  {\bibfnamefont {S.}~\bibnamefont {Kunii}},\ }\href {\doibase
  10.1103/PhysRevB.59.1808} {\bibfield  {journal} {\bibinfo  {journal} {Phys.
  Rev. B}\ }\textbf {\bibinfo {volume} {59}},\ \bibinfo {pages} {1808}
  (\bibinfo {year} {1999})}\BibitemShut {NoStop}%
\bibitem [{\citenamefont {Richtmyer}(1939)}]{Richtmyer_JAP1939}%
  \BibitemOpen
  \bibfield  {author} {\bibinfo {author} {\bibfnamefont {R.~D.}\ \bibnamefont
  {Richtmyer}},\ }\href {\doibase http://dx.doi.org/10.1063/1.1707320}
  {\bibfield  {journal} {\bibinfo  {journal} {J. Appl. Phys.}\ }\textbf
  {\bibinfo {volume} {10}},\ \bibinfo {pages} {391} (\bibinfo {year}
  {1939})}\BibitemShut {NoStop}%
\end{thebibliography}
%merlin.mbs apsrev4-1.bst 2010-07-25 4.21a (PWD, AO, DPC) hacked
%Control: key (0)
%Control: author (8) initials jnrlst
%Control: editor formatted (1) identically to author
%Control: production of article title (-1) disabled
%Control: page (0) single
%Control: year (1) truncated
%Control: production of eprint (0) enabled
%

\end{document}